\documentclass[twocolumn,floatfix,prl,showpacs]{revtex4}
\usepackage{graphicx}
\usepackage{amsmath}
\usepackage{bm}
\usepackage{hyperref}
\begin{document}
\title{Graphene-based electronic spin lenses}

\author{Ali G. Moghaddam and Malek Zareyan}

\affiliation{Institute for Advanced Studies in Basic Sciences
(IASBS), P.O. Box 45195-1159, Zanjan 45195, Iran}
\begin{abstract}
We theoretically demonstrate the capability of a
ferromagnetic-normal (FN) interface in graphene to focus an
electron-wave with a certain spin direction. The essential feature
is the negative refraction Klein tunneling, which is spin-resolved
when the exchange energy of F graphene exceeds its Fermi energy.
Exploiting this property, we propose a graphene NFN electronic
spin lens through which an unpolarized electronic beam can be
collimated with a finite spin-polarization. Our study reveals
that magnetic graphene has the potential to be the electronic counterpart of
the recently discovered photonic chiral meta-materials that
exhibit a negative refractive index for only one direction of the
circular polarization of the photon-wave.
\end{abstract}


\maketitle

There exists a close analogy between the propagation of photons
inside a photonic crystal and that of electrons in a solid state system
as a result of the wave-particle duality in quantum physics.
This analogy has been revealed by several counterpart effects in
the two progressing fields of photonics and solid state electron
optics \cite{e-optic}. Of particular interest in both fields has
been the focusing of a beam. In photonics, recent advances in
the fabrication of artificial meta-materials has provided the
ability to control the electromagnetic wave
flow inside matter almost completely. This is not possible in natural
materials. In particular, the realization of left-handed
meta-materials, which can have a negative refractive index
\cite{veselago,pendry-rev}, has shown exciting technological
promises such as perfect lenses \cite{perfect-lens} and electromagnetic
cloaking \cite{cloak}. On the other hand,
significant developments have been made in electron optics through the
fabrication of metallic and semiconducting nanostructures in which
the ballistic and phase-coherent transport of electrons make it
possible to observe electronic effects with photonic analogues
\cite{been89}. The idea of using quantum point contacts to focus
the electron wave in a two-dimensional electron gas subjected to a
magnetic field has already been experimentally achieved
\cite{gossard07}. The capability of graphene, a single
atomic layer of graphite, to become an electronic meta-material was
predicted recently \cite{falko07}. It was shown that an interface between
electron(n)-doped and hole(p)-doped regions in graphene can
focus an electron beam, which may lead to the realization of an
electronic Veselago's lens in analogy with the photonic
left-handed meta-materials.
\par
Despite the promising achievements in focusing the electron and
photon waves, until now, little attention has been paid
\cite{pershin-privman03,hoefer-silva-stiles07} to the polarization
degree of freedom of the focused beam. The most recent development
in photonics is the realization of the so called chiral
meta-materials \cite{chiral-meta1,chiral-meta2} in which the
degeneracy between the two circularly polarized waves is broken.
A strongly chiral meta-material may exhibit negative refraction
for one circularly polarized beam, while retaining positive
refraction for the other. Thus, the interface of
such a meta-material with an ordinary medium will focus only the
waves with a certain direction of the circular polarization, which
results in a circularly-polarized focusing of a linearly polarized
incident wave.
\par
\begin{figure}
\begin{centering}\vspace{-0.0cm}\par\end{centering}
\begin{centering}\includegraphics[width=7.cm]{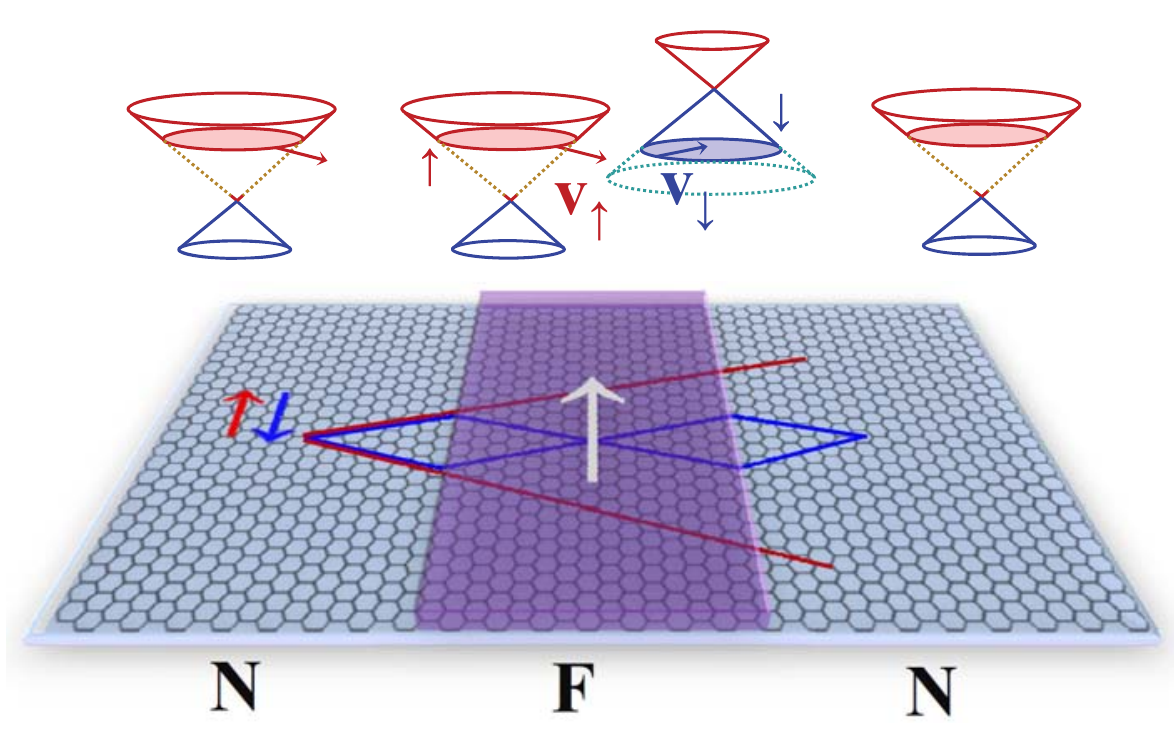}\par\end{centering}
\caption{\label{gzfig1}{\small (Color online) Schematic drawing of the ferromagnetic(F)
graphene spin lens in a normal(N) sheet. The configuration of the two spin sub-bands (being n
type or p type) is also shown,
when the electrostatic potential is set to $U_F=0$, $U_N=h$ in
F and N regions, respectively. The electrons are injected from a
nonmagnetic point source inside the left N region. A spin-up electron
beam (red lines) diverges, but a spin-down electron
beam (blue lines) undergoes negative refractions at the FN interfaces and
focuses in the right N region at the image point of the source. }}
\end{figure}
\par
In electron optics, however, the question of the possibility of
\textit{spin-polarized} focusing of an electron-wave has remained
unanswered. The aim of the present Letter is to
address this question by introducing a model based on magnetic
graphene.  We show that a weakly doped ferromagnetic (F) graphene can
be the electronic counterpart of photonic chiral meta-materials, in the
sense that it can be used for focusing electrons with a certain
spin direction. Based on this finding, we propose an electronic spin
lens, shown schematically in Fig. \ref{gzfig1}, through which an
unpolarized incident electron-wave can be focused into an image point with
a finite spin-polarization. The spin-polarization of the image is
directed anti-parallel to the magnetization vector of F. Such a
possibility for the realization of a focused spin accumulation with
a tunable direction can also be of great interest in the field of
spintronics \cite{SpnReview}.
\par
The potential of graphene to be used for electron focusing is suggested
by its unique zero-gap semiconducting electronic band structure
\cite{nov05,zha05,castro-rmp}. Its conically shaped conduction and
valence bands touch each other at the corners of hexagonal first
Brillouin zone, known as Dirac points. The carrier type and its
density can be tuned by means of electrical gates or by doping the
underlying substrates. At low energies, the quasi-particles are
described by the massless Dirac Hamiltonian, $\hat{H}_D=v_F {\bf
p}\cdot\hat{\bm \sigma}$, with Fermi velocity $v_F$, momentum
${\bf p}$, and Pauli matrices $\hat{\bm
\sigma}=(\sigma_x,\sigma_y)$ defined in pseudo-spin space to
characterize the two trigonal sub-lattices of the hexagonal
structure of graphene. The linear dispersion, together with
the pseudo-spin aspect, give the carriers a pseudo-relativistic
chiral nature with electrons and holes having different chiralities ${\bf
p}\cdot\hat{\bm \sigma}/p=\pm1$. An important manifestation of the
chirality is the so called Klein tunneling which is a negative
refraction process through a p-n junction \cite{falko06,katsnelson,goldhaber}.
\par
An interesting consequence of the specific band structure described above
is that in a ferromagnetic graphene with an exchange
potential exceeding its Fermi energy, the Fermi level for the spin-up
and the spin-down carriers lies in the conduction and the valence
spin-subbands, respectively \cite{neto05,zmg08}. This means that
the opposite-spin carriers are of different types, electron-like and
hole-like, and hence, have opposite chiralities. We show that the
interface between such \textit{spin-chiral} materials and nonmagnetic graphene
(with the same type of carriers in the two spin-subbands)
exhibits negative refraction for electrons with
a certain spin-direction, while retaining positive refraction
for electrons with an opposite spin direction. We demonstrate that
spin-resolving the sign of the electronic refractive index in this manner can lead to
the realization of a graphene normal-ferromagnetic-normal (NFN) spin lens. 
\par
Our model consists of a spin-chiral F stripe of width $w$
inside an N graphene sheet as shown in Fig. \ref{gzfig1}. Such an
F region can be produced by using, in part, an insulating ferromagnetic
substrate. Alternatively, F metals or added magnetic impurities on
top of a graphene sheet can induce an exchange potential
\cite{neto08,wees}. Intrinsic ferromagnetic correlations are also
predicted to exist in graphene sheets \cite{neto05} and nanoribbons
with zigzag edges \cite{louie06} under certain conditions. To study
the focusing effect, we use the single-electron Green's function method.
The Hamiltonian for a spin-$s (=\pm)$ electron in one of the valleys
is given by
\begin{equation}
\hat{H}^{0}_{s}=\hat{H}_D-sh({\bf r})-U({\bf r}),
\label{hamiltonian}
\end{equation}
where $h({\bf r})$ and $U({\bf r})$ are the exchange and
the electrostatic potential, respectively, and are functions
of the 2D position vector ${\bf r}$. We model a nonmagnetic
electronic point source at the position ${\bf r}_0\equiv(x_0,y_0)$
in the left N region as the perturbation potential
$\hat{V}_s=\hat{V}_0\delta({\bf r}-{\bf r}_0)$, with strength
$\hat{V}_0$. The total Hamiltonian then, becomes
$\hat{H}_{s}=\hat{H}^{0}_{s}+\hat{V}_s$. The local density of
states (LDOS) of spin-$s$ electrons can be calculated using the
relation $n_{s}(\varepsilon,{\bf r})=-(1/\pi){\rm Im} \, {\rm Tr}
\,\hat{G}_{s}({\bf r}|{\bf r})$ in which the retarded
Green's function is defined as
\begin{eqnarray}
\hat{G}_{s}({\bf r}|{\bf
r'})=\lim_{\eta\rightarrow0^{+}}\langle{\bf r}
|(\varepsilon+i\eta-\hat{H}_{s})^{-1}|{\bf r'}\rangle,
\end{eqnarray}
with ${\rm Tr}$ denoting the trace over the space of the
pseudo-spin. Using the Dyson expansion, the change of the LDOS
induced by the perturbation up to the first order in $\hat{V}_{0}$
can be calculated from the equation
\begin{eqnarray}
\delta n_{s}({\bf r})&=&-\frac{1}{\pi}{\rm Im} \, {\rm Tr}
\,[\hat{G}_{s}^{0}({\bf r}|{\bf
r}_0)\hat{V}_{0}\hat{G}_{s}^{0}({\bf r}_0|{\bf r})],
\label{deltans}
\end{eqnarray}
in which the unperturbed Green's function $G^{0}_{s}$ satisfies the
relation $[\varepsilon-\hat{H}^{0}_{s}({\bf r})]\hat{G}_{s}^{0}({\bf r}|{\bf
r'})=\delta({\bf r}-{\bf r'}). $
\par
A voltage $V$ applied to the source point can induce a current
in the left N region, which we found to be spin-polarized.
The current-density for spin-$s$ electrons is obtained from
\begin{eqnarray}
{\bf i}_{s}({\bf r})/eV= ev_{F}{\rm Im} \, {\rm
Tr}[\hat{\bm \sigma} \hat{G}_{s}^{0}({\bf r}|{\bf
r}_0)i\Gamma\hat{G}_{s}^{0\dag}({\bf r}_0|{\bf r})],
\label{deltais}
\end{eqnarray}
where the level broadening function $\Gamma$ is the measure of the
tunneling rate between the source lead and the N graphene sheet.
\par
Assuming that the potentials $U$ and $h$ vary only along the $x$
direction, we can use the Fourier transformation
$\hat{G}^{0}_{s}({\bf r}|{\bf r'})=\int d
p_{y}\exp[ip_{y}(y-y')]\hat {g}_{s,p_{y}}(x|x')$. The new Green's
function $\hat{g}$ satisfies the one dimensional evolution-like
equation (as a function of the position $x$ instead of the time),
\begin{eqnarray}
[iv_{F}\partial_{x}-\hat{\cal
L}_{s,p_y}(x)]\hat{g}_{s,p_y}(x|x')=\sigma_x\delta(x-x'),
\label{evlike}
\end{eqnarray}
with a non-Hermitian Hamiltonian $\hat{\cal L}_{s,p_y}(x)=-[U(x)+sh(x)+\varepsilon]\hat{\sigma}_x+iv_{F}p_{y}\hat{\sigma}_z.
$
In principle Eq. (\ref{evlike}) together with Eqs. (\ref{deltans}), (\ref{deltais})
can be solved numerically to obtain the spin-resolved variations of the LDOS and
the current density for the given profiles of $h({\bf r})$ and $U({\bf r})$.
\par
Before we proceed with the full quantum mechanical
calculation, we may apply the adiabatic approximation to the
non-Hermitian Hamiltonian $\hat{\cal L}$ in Eq. (\ref{evlike}),
which is valid when the variation of the potentials is slow
on the scale of the Fermi wave-length in N and F. In this way
the semi-classical expression of $\hat{G}_{s}^{0}$ is obtained,
from which we deduce that the semiclassical trajectory of a
spin-$s$ electron in N and F regions consists of straight lines
given by the relations
\begin{eqnarray}
y-y_0+x_0\tan\theta_N=~~~~~~~~~~~~~~~~~~~~~~~~~~~~~~~~~~~~~~~\nonumber\\
\left\{%
\begin{array}{ll}
x\tan\theta_{N}   , & x<x_L\hbox{;} \\
    (x-x_L)\tan\theta_{Fs}+x_L\tan\theta_{N}, &  x_L<x<x_R\hbox{;} \\
    (x-w)\tan\theta_{N}+w\tan\theta_{Fs}, & x>x_R\hbox{.} \\
\end{array}%
\right. ~~~~ \label{rays}
\end{eqnarray}
Here $\theta_{N}=\arcsin(p_y/\mu_{N})$ and $\theta_{Fs}=\arcsin(p_y/\mu_{Fs})$
are the propagation angles (measured from the normal to FN interfaces) and
$\mu_{Fs}$ and $\mu_{N}$ are the electrochemical potentials for a spin-$s$ electron
inside F and N regions, respectively; $x_{L(R)}$ indicates the locations of the
left (right) interface.
\par
From the relations (\ref{rays}) we find that the focusing can
occur for spin-down $s=-$ $(h>0)$ electrons provided that
$\mu_{F-}$ and $\mu_N$ have opposite signs. In this case, the angle
$\theta_{F-}$ undergoes a change of sign at both FN interfaces,
indicating that the NFN structure operates as a \textit{spin}
n-p-n structure for spin-down electrons. Eq. (\ref{rays}) also gives
the location of the two focuses inside F and the right N region as
$x_{F}-x_L=(x_0-x_L)\tan\theta_N/\tan\theta_{F-}$ and
$x_{N}-x_0=w(1-\tan\theta_{F-}/\tan\theta_N)$, respectively. We
note that in general, the location of the focal point depends on
the transverse momentum $p_y$ of the incident electron, which
could lead to the appearance of many focusing points. This problem
can be solved if we set a symmetric spin p-n potential profile
at the interfaces by having $\mu_N=-\mu_{F-}$, which results in a
unique, profound focus at $x_F-x_L=-(x_0-x_L)$ and
$x_{N}-x_0=2w$. We note that even with a symmetric profile at the
interfaces, only the electrons close to the Fermi level are
focused effectively, which shows the effectiveness of the focusing
at low temperatures. From Eq. (\ref{rays}) we have estimated that at
a finite temperature $T$ the focal point will spread along the x
direction over a length of the order
$(k_BT/\mu_N)L$, where L is the distance of the source from the
left FN interface ($k_B$ is the Boltzmann constant). To have a
profound spin-lensing, this length should be much smaller than $L$,
which gives a rough estimate of $T<\mu_N/k_B$.
With $\mu_N\sim 10-100~meV$ in graphene sheets,
a temperature lower than $100 K$ is thus required.
\par
On the other hand, to have a significant spin-polarization at the
focal points, the spin-up electrons have to remain unfocused.
This is achieved by assigning the same sign to both $\mu_{F+}$ and
$\mu_N$, which means that the spin-up electrons
remain at the same subband (valence or conduction) throughout
the whole structure. Let us consider two special cases of
$\mu_{F+}=\mu_N$ and $\mu_{F+}=0$. In
the first case, a spin-up electron does not feel any potential
change and thus propagates divergently through the system.
In the second case, the density of states of spin-up
electrons vanishes in F. This implies that a spin-up electron
cannot propagate into the F region, but rather tunnels through
evanescent modes, which have a small contribution to the variation
of LDOS. These two specific cases correspond to the potential sets of
$U_F=0$, $U_N=h$ and $U_F=-h$, $U_N=2h$, respectively.
\par
\begin{figure}
\begin{centering}\vspace{-0.1cm}\par\end{centering}
\begin{centering}\includegraphics[width=8.5cm]{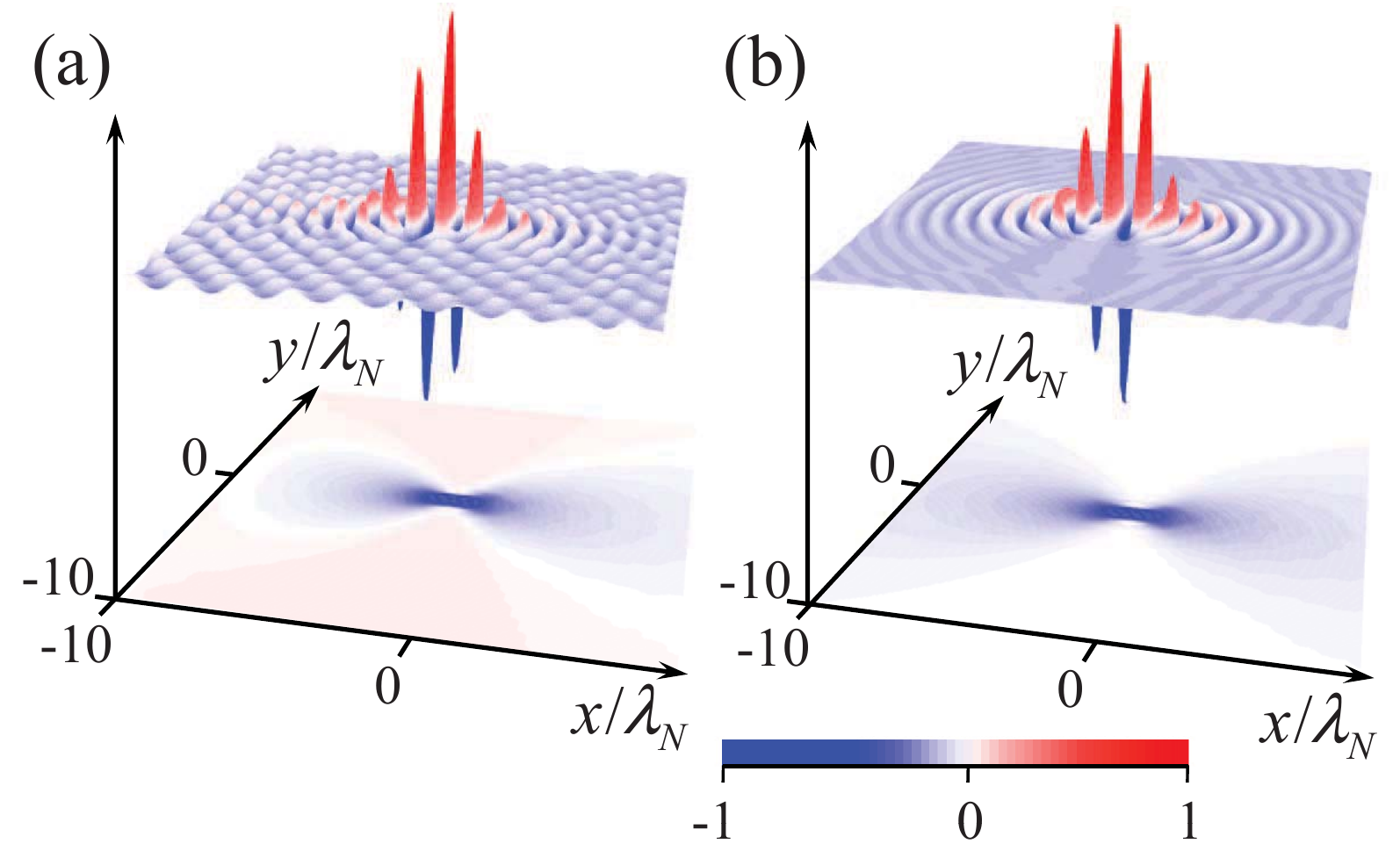}
\par\end{centering}
\caption{\label{fig2}{\small (Color online) Spin local density of states (upper
plots) and the amplitude of the spin-current density (lower plots)
around the image point in the left N region (both in arbitrary units), when
a nonmagnetic point source is located in the right N region. The potential is
set to (a) $U_F=0$, $U_N=h$ and (b) $U_F=-h$, $U_N=2h$.  Both quantities
have peaks around the image point of the source, with spin
LDOS showing Friedel-like oscillations around its peak.}} \label{gzfig2}
\end{figure}
\par
Figure \ref{gzfig2} shows the result of our quantum calculation for
the spin LDOS, defined as $\delta n_{+}-\delta n_{-}$, inside the
right N region when a point nonmagnetic perturbation is located in
the left N region and for the two sets of potential (a) $U_F=0$, $U_N=h$
and (b) $U_F=-h$, $U_N=2h$ described above.
The distribution of the amplitude of the spin-current
density, $|{\bf i}_{+}|-|{\bf i}_{-}|$, is also shown when a
voltage $V$ is applied to the source point. We have assumed
that the potential varies abruptly at the FN interfaces.
The spin LDOS shows a peak at the image point of the perturbation
with Friedel-like oscillations whose period is of the order of the Fermi
wavelength $\lambda _N=\hbar/\mu_N$ in N.
The difference between the two sets of potential is visible at
points far from the image point. We note that
focusing electrons by an NFN graphene creates a mirage that
replicates LDOS oscillations which, unlike
the original perturbation, are spin-polarized and mimic the effect
of a magnetic perturbation at the image point. Thus, the NFN structure
produces a magnetic image from a nonmagnetic point source.
\par
\begin{figure}
\begin{centering}\vspace{-0.01cm}\par\end{centering}
\begin{centering}\includegraphics[width=8.5cm]{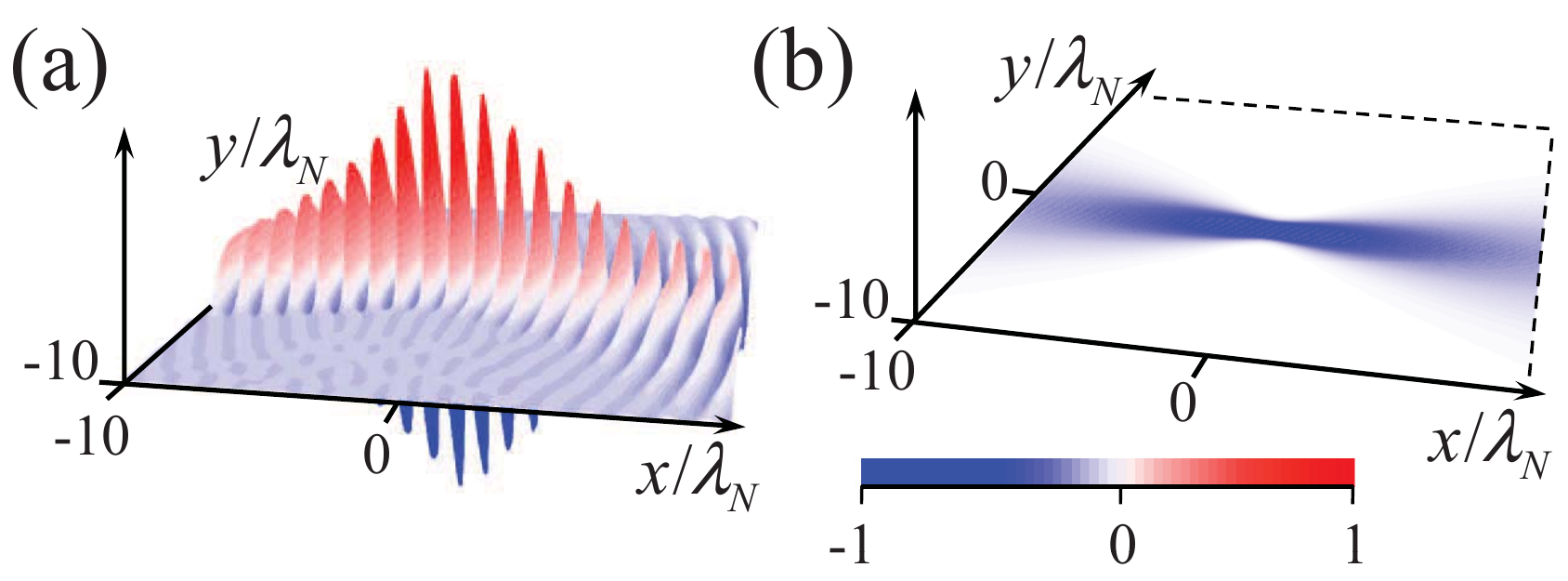}
\par\end{centering}
\caption{\label{fig3}{\small (Color online) Effect of the smooth variation of the
potential at FN-interfaces on the spin image of Fig. \ref{gzfig2}b.
The potential is assumed to vary linearly across
the interfaces from $U_F=-h$ to $U_N=2h$ over a distance $\Delta
x=\lambda _{N}$. Compared to the abrupt interfaces, the peaks of
spin-LDOS (a) and of the amplitude of the spin-current density (b)
are broadened.
 }}
 \label{gzfig3}
\end{figure}

We have also investigated the effect of the smooth variation of the potential
at the interfaces on spin lensing. The result is shown in Fig. \ref{gzfig3},
in which the spin LDOS (Fig. \ref{gzfig3}a) and the distribution of the amplitude
of the spin current-density (Fig. \ref{gzfig3}b) are plotted for the potential
set of Fig. \ref{gzfig2}b, but for a finite thickness of the interfaces
$\Delta x=\lambda _{N}$, over which $U(x)$ varies linearly from $U_F=-h$
to $U_N=2h$. Compared to the case of sharp interfaces, the peaks of
the spin-LDOS and of the amplitude of the spin-current density
are broadened. Therefore, introducing a smooth variation of the potential
at the interfaces leads to the weakening of spin lensing.
We note that the potential variation length is restricted in graphene
because of the screening effect \cite{fogler}. This, together with the low carrier
densities and large Fermi wavelengths of graphene, make it possible to
envisage contacts smaller than the Fermi wavelength.
Spin lensing can be observed experimentally by spin-polarized
scanning tunneling microscopy \cite{bode03} of the N graphene
region around the focal point, which can image the variation of
the spin LDOS with a resolution of the order of a few $nm$.
\par
Regarding the validity of the independent valleys model \cite{castro-rmp}
of the Hamiltonian(\ref{hamiltonian}), it is well known \cite{akhbeen08}
that unlike a p-n contact in graphene nanoribbons with zigzag edges for which
the intrinsic inter-valley mixing is strong no matter how smooth the potential
variation might be, in the wide contact geometry of our model, the inter-valley
scattering becomes effective only for very abrupt contacts of length in
the order of an atomic lattice constant $a \sim 1 {\AA}$.
In graphene, this length scale is much shorter than $\lambda_N$,
which is typically a few hundred $nm$. We have found that spin lensing is effective
for a contact of lengths up to $\lambda_N$.
Thus, our assumption that the inter-valley mixing is negligible
is well justified for contacts with a length smaller than $\lambda_N$, but
much larger than $a$.
\par
In conclusion, we have proposed a solid state electronic spin lens
based on a ferromagnetic graphene which has an exchange
potential higher than its Fermi energy. The key property is that
an interface between such a spin-chiral F and an N
graphene region exhibits an electronic refractive index which has
different signs for electrons with different spin-directions. We
have shown that in a corresponding NFN structure, a point-like
nonmagnetic source in one N region produces an image in the other N
region which is a point spin accumulation with associated
Friedel-like oscillations of spin LDOS.

\end{document}